\documentclass[conference]{IEEEtran}
\usepackage{array}
\usepackage{multirow}% http://fctan.org/pkg/multirow
\usepackage[noadjust]{cite}
\usepackage{algorithm}
\usepackage{algorithmic}
\usepackage{amssymb}
\usepackage{bm}
\usepackage{amsmath}
\usepackage{cases} 
\usepackage{multirow}
\usepackage{booktabs}
\usepackage[caption=false,font=footnotesize]{subfig}
\usepackage{siunitx}
\usepackage{tabularx}
\usepackage{makecell}
\usepackage[table,xcdraw]{xcolor}
\ifCLASSINFOpdf
  \usepackage[pdftex]{graphicx}
  \graphicspath{{./}{./}}
  \DeclareGraphicsExtensions{.pdf,.jpeg,.png}
\else
  \usepackage[dvips]{graphicx}
  \graphicspath{{./}}
  \DeclareGraphicsExtensions{.eps}
\fi
\begin{document}
\title{TC: Throughput Centric Successive Cancellation Decoder Hardware Implementation for Polar Codes   
\vspace{-0.5em}}

\author{\IEEEauthorblockN{Tiben Che, Jingwei Xu and Gwan Choi}
\IEEEauthorblockA{Department of Electrical and Computer Engineering\\
Texas A\&M University, College Station, Texas 77840\\
Email: $\lbrace$ctb47321, xujw07, gchoi$\rbrace$@tamu.edu}}


%Telephone: (800) 555--1212, Fax: (888) 555--1212}


\maketitle

\begin{abstract}
%\boldmath
This paper presents a hardware architecture of fast simplified successive cancellation (fast-SSC) algorithm for polar codes, which significantly reduces the decoding latency and dramatically increases the throughput. 
Algorithmically, fast-SSC algorithm suffers from the fact that its decoder scheduling and the consequent architecture depends on the code rate; this is a challenge for rate-compatible system. However, by exploiting the homogeneousness between the decoding processes of fast constituent polar codes and regular polar codes, the presented design is compatible with any rate. 
The scheduling plan and the intendedly designed process core are also described. 
Results show that, compared with the state-of-art decoder, proposed design can achieve at least $60\%$ latency reduction for the codes with length $N=1024$. By using $Nangate~FreePDK~45nm$ process, proposed design can reach throughput up to $5.81~Gbps$ and $2.01~Gbps$ for $(1024,870)$ and $(1024,512)$ polar code, respectively.      

\end{abstract}


% \fi
\IEEEpeerreviewmaketitle

\section{Introduction}
Recently, polar codes~\cite{arikan2009channel} have received significant attention due to its capability to achieve the capacity of binary-input memoryless symmetric channels with low-complexity encoding and decoding schemes.
Successive cancellation (SC)~\cite{arikan2009channel}, list successive cancellation (List-SC)~\cite{tal2011list} and belief propagation (BP)~\cite{xu2015xj} are the three most common proposed decoding schemes. 
Among these, SC decoder is the most promising for practical hardware implementation since its low $O(NlogN)$ complexity, where $N$ is the length of the code. Thus, many relevant hardware designs are proposed~\cite{leroux2011hardware}~\cite{leroux2013semi}~\cite{mishra2012successive}.
%The tree architecture and the line architecture of SC decoder are proposed in~\cite{leroux2011hardware}, with synthesis results using $TMSC$ $65nm$ process. 
%A more efficient semi-parallel SC decoder design is proposed for both tree architecture and line architecture~\cite{leroux2013semi}.
%An ASIC design of SC decoder in $180nm$ CMOS process is described in~\cite{mishra2012successive}.

However, algorithmically, SC decoder suffers from high latency. Typically, for conventional SC decoder, its latency ($2N-2$) increases linearly with respect to the code length.
This is a significant challenge since polar codes work well only at very long code lengths. 
A lot of works have been done to reduce the latency of SC decoder from both hardware and algorithm aspects. 
In~\cite{zhang2012reduced}, a pre-computation  method is used to reduce decoding latency from $2N-2$ to $N-1$.
In~\cite{yuan2014low}, three approaches, the dedicated 2-bit decoder for the last stage of SC decoding, overlapped-scheduling and look-ahead techniques are applied, which eventually results in a $3N/4-1$ latency. 
In~\cite{alamdar2011simplified} and ~\cite{sarkis2014fast}, by observing the tree architecture of SC decoding, certain patterns of constituent codes are found. 
These constituent codes can feed back the hard decision information immediately without traversal, which can significantly reduce the latency of decoding some polar codes with a given architecture. 
This approach is refer to as fast-SSC decoder. Moreover, a processors-array based structure for FPGA implementation is also proposed in~\cite{sarkis2014fast}.

In this paper, a novel low latency hardware architecture of polar code decoding using fast-SSC algorithm is presented. 
Although fast-SSC algorithm naturally lacks flexibility for multiple rates, the proposed design overcomes this disadvantage by utilizing the similarity between the decoding processes of fast constituent polar codes and regular polar codes. Corresponding scheduling plan is presented in this paper. We also provide the design details of the $processing~unit$ (PU) which is compatible with both regular polar code and constituent polar code. The comparisons with other commonly discussed SC decoders are given. For example, Compared with the 2b-SC-Precomputation decoder, the fastest ASIC design of SC decoder to best of our knowledge, the proposed design can achieve at least $60\%$ latency reduction for polar code with length $N=1024$. 
%For example, a reduction of $79.66\%$ at the rate of $0.85$ and $65.31\%$ at the rate of $0.5$. 
The analysis of latency reduction with respect to code rates is also presented. It shows proposed architecture can yield a significant latency reduction especially at high code rate (code rate $>$ 0.8). This is very promising for modern communication or data storage systems where high rate codes are desired. Synthesis results using $Nangate~FreePDK~45nm$ process shows the proposed design can reach throughput of up to $5.81~Gbps$ and $2.01~Gbps$ for $(1024,870)$ and $(1024,512)$ polar codes, respectively. 

This paper is organized as follows. The relative background are reviewed in section~\ref{Background}. Then, the hardware implementation of proposed system is described in section~~\ref{Hardware Implementation}. After that, the synthesis results and relevant comparisons are discussed in section~~\ref{Hardware Analysis of Comparison}. Finally, the conclusion is in section~\ref{Conclusion}.

\section{Background}
\label{Background}
\subsection{Polar Code and Tree analysis of SC Decoding}
\label{Polar Code}

%\begin{figure}[!t]
%\centering
%\includegraphics[width=3in]{encoder.eps}
%\caption{Polar code encoder for codelength 8}
%\label{encoder}
%\end{figure}

\begin{figure}[!t]
\centering
\subfloat[]{\includegraphics[width=1.5in]{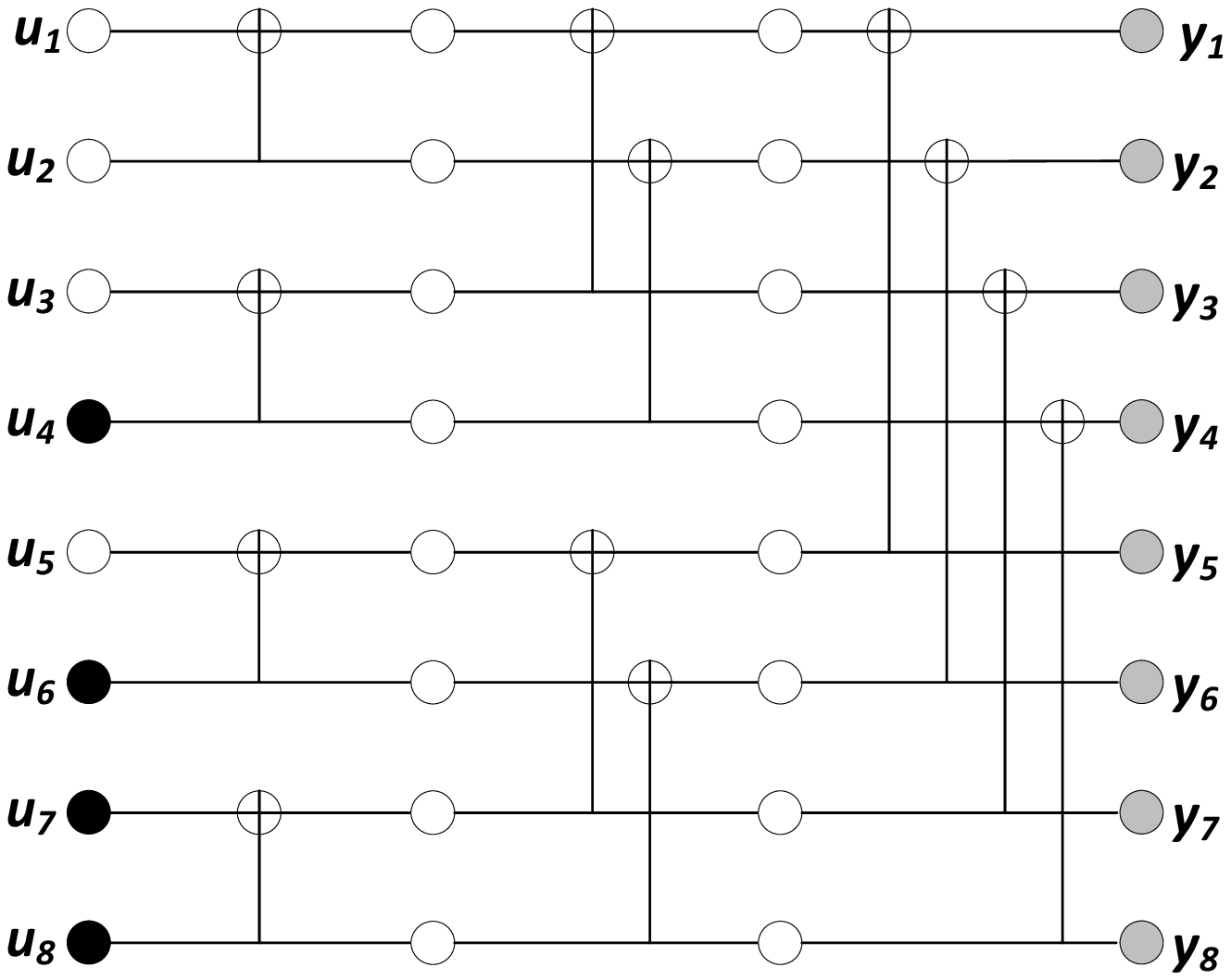}\label{encoder}}
\hfil
\subfloat[]{\includegraphics[width=0.75in]{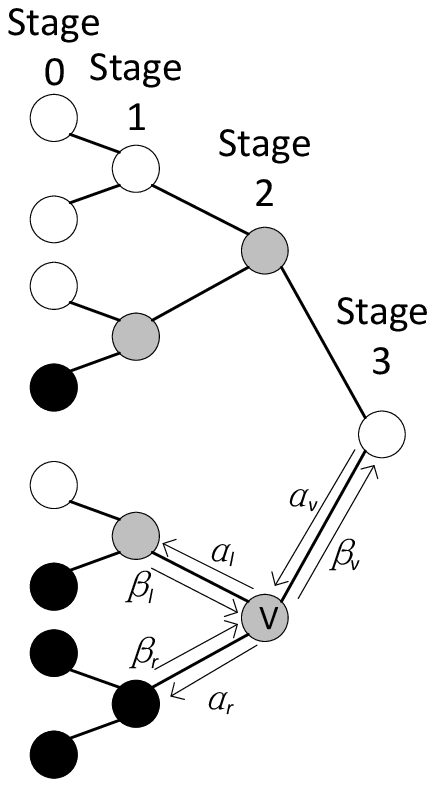}\label{tree}}
\hfil
\caption{\protect\subref{encoder} Encoder of $(8,4)$ polar code, \protect\subref{tree} Tree presentation of $(8,4)$ SC decoder}
\label{polar_code_introduction}
\end{figure}  

As described in \cite{arikan2009channel}, a polar code is constructed by exploiting channel polarization.
%Fig.\ref{encoder} shows an implementation of polar encoder when codelength is 8. 
Mathematically, polar codes are linear block codes of length $N = 2^n$. The transmitted codeword ${\bm{x}}\triangleq {(x_1,x_2,\cdots,x_N)}$ is computed by $\bm{x}=\bm{u}\bm{G}$ where $\bm{G=F^{\otimes m}}$, and $\bm{F^{\otimes m}}$ is the $m$-th Kronecker power of 
$\bm{F} = 
\begin{bmatrix}
1&0\\
1&1
\end{bmatrix}
$. 
Each row of $G$ is corresponding to an equivalent polarizing channel. For an $(N,k)$ polar code, $k$ bits that carry source information in $\bm{u}$ are transmitted using the most reliable channels. These are refer to information bits. While the rest $N-k$ bits, called frozen bits, are set to zeros and are placed at the least reliable channels. Determining the location of the information and frozen bits depends on the channel model and the channel quality is investigated in~\cite{tal2013construct}. Fig.~\ref{encoder} shows an example of $(8,4)$ polar code encoder, where the black and white nodes stand for the information bits and frozen bits, respectively.   

Polar codes can be decoded by recursively applying successive cancellation to estimate $\hat{u}_i$ using the channel output $y_{0}^{N-1}$ and the previously estimated bits $\hat{u}_{0}^{i-1}$. 
This approach is naturally represented by a binary tree whose each node corresponds to a constituent code. The number of bits in one constituent node in stage $m (m = 0,1,2...)$ $N^{m}$ is equal to $2^m$. Fig.~\ref{tree} shows an example of $(8,4)$ polar code. 
$\bm{\alpha}$ stands for the soft reliability value, typically is log-likelihood ratio (LLR), and $\bm{\beta}$ stands for the hard decision.      
$\bm{\alpha_l}$ and $\bm{\alpha_r}$ are the message passing from parent node to left and right child, and can be computed according to Eq.~(\ref{left_child}) and Eq.~(\ref{right_child}), respectively. 
\begin{equation}
\begin{aligned}
\alpha_{l}[i] &= f(\alpha_{v}[i],\alpha_{v}[i+N^{m}/2])\\ 
&= sign(\alpha_{v}[i])sign(\alpha_{v}[i+N^{m}/2])\\
&\cdot~min(|\alpha_{v}[i]|,|\alpha_{v}[i+N^{m}/2]|)
\label{left_child}
\end{aligned}  
\end{equation}
\begin{equation}
\begin{aligned}
\alpha_{r}[i] &= g(\beta_{l}[i-N^{m}/2],\alpha_{v}[i],\alpha_{v}[i-N^{m}/2]) \\
&= (-1)^{\beta_{l}[i-N^{m}/2]}\cdot\alpha_{v}[i-N^{m}/2]+\alpha_{v}[i]
\label{right_child}
\end{aligned}
\end{equation}
At stage 0, $\beta_v$ of a frozen node is always zero, and for information bit its value is calculated by threshold detection of the soft reliability according to 
 \begin{equation}\label{hard_decision}
  \beta_v=h(\alpha_v)=  
  \left
  \{
   \begin{array}{c}
   0,~if~\alpha_v \geqslant 0  \\
   1,~otherwise  \\
   \end{array}
  \right.
 \end{equation}  
At intermediate stages, $\bm{\beta_v}$ can be recursively calculated by
 \begin{equation}\label{feedback}
  \beta_{v}[i] =  
  \left
  \{
   \begin{array}{ll}
   \beta_{l}[i] \oplus \beta_{r}[i]~if~i\leq~N^{m}/2 \\
   \beta_{r}[i-N^{m}/2]~otherwise  \\
   \end{array}
  \right.
 \end{equation}      
%$\bm{\beta_l}$ and $\bm{\beta_v}$ are the hard decisions of each constituent code. $\bm{\beta_v}$ is the hard information needed to feedback to parent stage, and it can be computed by $\bm{\beta_l}$ and $\bm{\beta_v}$ via Eq.~(\ref{feedback}). 

%\begin{figure}[]
%\centering
%\subfloat[]{\includegraphics[width=1in]{n0n1.eps}\label{n0n1}}
%\hfil
%\subfloat[]{\includegraphics[width=1.04in]{nsnr.eps}\label{nsnr}}
%\caption{\protect\subref{tree} Tree presentation of 8-bit SC decoder, \protect\subref{n0n1} An example of $\mathcal{N}^0$ and $\mathcal{N}^1$ in a 8-bit SC tree, and \protect\subref{nsnr} An example of $\mathcal{N}^{SPC}$ and $\mathcal{N}^{REP}$}
%\label{3tree}
%\end{figure}

\subsection{Fast-SSC Algorithm}
\label{Fast SC Algorithm}

The main idea of fast-SSC algorithm is illustrated in~\cite{zhang2012reduced},~\cite{alamdar2011simplified} and~\cite{sarkis2014fast}. By finding some certain pattern constituent polar codes, the hard decision $\bm{\beta_v}$ of each constituent node can be determined immediately, without traversing the entire subtree, once the constituent polar code is activated. For a length $N$ constituent code in non-systematic polar codes, $\hat{\bm{u}}_N$ is calculated by $\hat{\bm{u}}_N = \bm{\beta_{vN}} \cdot G_N$, where $G_N$ is the generator matrix for length $N$ polar code. We adopt four kinds of constituent polar codes in our design.
These are $\mathcal{N}^0$, $\mathcal{N}^1$, $\mathcal{N}^{SPC}$ and $\mathcal{N}^{REP}$, which are called fast constituent polar codes.

$\mathcal{N}^0$ and $\mathcal{N}^1$ are refer to those constituent codes which only contain frozen bits or information bits, respectively.
For $\mathcal{N}^0$ codes, we can set $\bm{\beta_v}$ to $0$ immediately. 
For $\mathcal{N}^1$ node, $\bm{\beta_v}$ can be directly decided via threshold detection Eq.~(\ref{hard_decision}). 
$\mathcal{N}^{SPC}$ and $\mathcal{N}^{REP}$ are two kinds constituent codes containing both frozen bits and information bits. In a length $N$ $\mathcal{N}^{SPC}$ codes, only the first bit is frozen. It renders the constituent codes as a rate $(N-1)/N$ single parity check (SPC) code. This code can be decoded by doing parity check with the least reliable bit which has the minimum absolute value of LLR. 
First, get the hard decision $HD_{v}$ of $\bm{\beta_v}$ via threshold detection. 
Then, calculated the parity by  
\begin{equation}\label{HD_SPC}
parity = \sum_{i=1}^{N^m} \oplus HD_{v}[i].
\end{equation}   
and, find the index of the least reliable bit via 
\begin{equation}\label{min_SPC}
j= arg \min_i|\alpha_{v}[i]|.
\end{equation}     
Eventually, $\bm{\beta_v}$ is decided by 
 \begin{equation}\label{beta_SPC}
  \beta_{v}[i]=  
  \left
  \{
   \begin{array}{ll}
   HD_{v}[i]\oplus parity,~when~i=j   \\
   HD_{v}[i],~otherwise  \\
   \end{array}
  \right.
 \end{equation}
In a length $N$ $\mathcal{N}^{SPC}$ codes, only the last bit is information bit. In this case, all the $\beta_{v}[i]$ should be the same and are reflections of the information contained in the only one information bit. Thus, the decoding algorithm starts by summing all input LLRs and $\bm{\beta_v}$ is calculated as 
 \begin{equation}\label{beta_REP}
  \beta_{v}[i]=  
  \left
  \{
   \begin{array}{ll}
   0,~when~\sum\alpha_{v}[i] \geqslant 0;   \\
   1,~otherwise  \\
   \end{array}
  \right.
 \end{equation}

\begin{figure}[]
\centering
\subfloat[]{\includegraphics[width=1in]{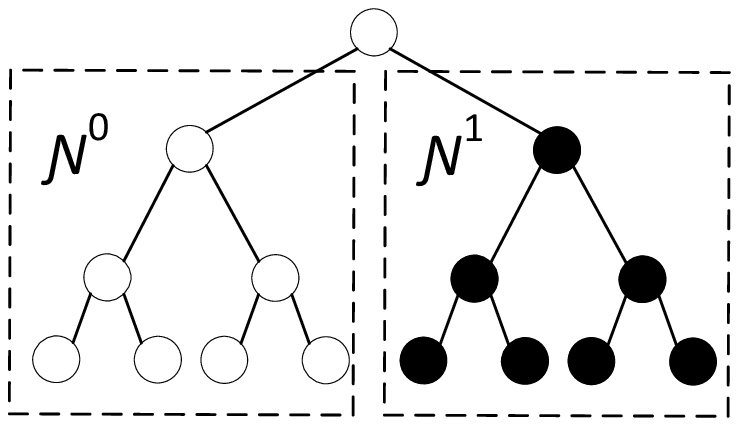}\label{n0n1}}
\hfil
\subfloat[]{\includegraphics[width=1in]{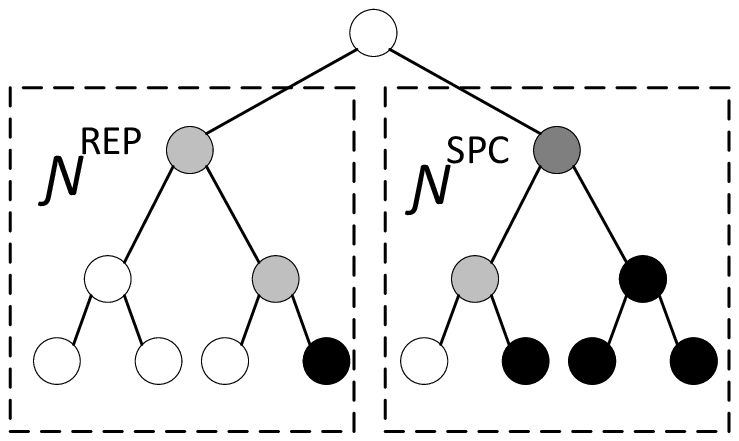}\label{nsnr}}
\caption{\protect\subref{n0n1} An example of $\mathcal{N}^0$ and $\mathcal{N}^1$ in a 8-bit polar code tree, and \protect\subref{nsnr} An example of $\mathcal{N}^{SPC}$ and $\mathcal{N}^{REP}$ in a 8-bit polar code tree} 
\label{2tree}
\end{figure}
    
Fig.~\ref{2tree} gives the examples of tree presentations of these four kinds constituent polar codes. 
%It can easily tell that the distribution of constituent is decided by the code rate. 
%Black circle and white circle stand for information bit and frozen bit, respectively. Actually, in~\cite{sarkis2014fast}, there is one more constituent polar code $\mathcal{N}^r$ has been introduced.  $\mathcal{N}^r$ contains multiple frozen bits and information bits, and can be decoded by exhaustive-search maximum-likelihood (ML) algorithm. However, ML detection is not efficient in hardware, thus we do not adopt $\mathcal{N}^r$ in our proposed design. $\mathcal{N}^0$, $\mathcal{N}^1$, $\mathcal{N}^{SPC}$ and $\mathcal{N}^{REP}$ constituent codes are prevalent in long length polar code~\cite{sarkis2014fast}, which allows us to gain a significant throughput improvement. 
%This can be shown in table~\ref{codes_sizes} where lists number of fast constituent codes with different sizes in a (1024, 512) polar code with rate of $0.5$.
  
% \usepackage{multirow}

%\begin{algorithm}[!htbp]
%\small
%\caption{TMR Control Algorithm}
%\label{alg_para_sub}
%\begin{algorithmic}[1]
%%\STATE Initialize $\pmb{a}^{0}=0, \pmb{v}=\pmb{u}, k=0, S=\emptyset;$
%\WHILE{the $1st$ and $2nd$ data have not arrived}
%	\STATE Wait;
%\ENDWHILE\
%\IF {the $1st$ arrived data = the $2nd$ arrived data}
%    \STATE output either of them;
%    \STATE set the acknowledgment signal;
%\ELSE
%	\WHILE{the $3rd$ data have not arrived}
%		\STATE Wait;
%	\ENDWHILE\
%	\STATE output the $3rd$ arrived data
%	\STATE set the acknowledgment signal;
%\ENDIF

%\end{algorithmic}
%\end{algorithm}

\section{Hardware Implementation}
\label{Hardware Implementation}

%In this section we introduce the hardware implementation of fast SC decoder.  A scheduling scheme is proposed in our design to deal with these 4 constituent polar codes and regular polar codes in one frame. We also can observe that the distribution of constituent polar code depends on the length and rate of the code. In order to compatible with multiple rate, a general re-configurable architecture is proposed. Besides, design details of  a special processing unit which can compatible with both regular polar code or constituent polar code by sharing adder and comparator is also presented in this section.   
In this section, a novel hardware implementation of fast-SSC decoder is presented. 
For a polar code with a given length, different code rate yields different distribution of constituent polar codes. 
A thoughtfully-composed architecture should have the capability and flexibility to deal with different rates.  
By exploiting the homogeneousness between the decoding processes of fast constituent polar codes and regular polar codes, our design supports a variety of rates. 
The scheduling scheme based on the proposed architecture is also discussed. 
Additionally, we develop an approach for sharing and reusing computational elements to achieve higher hardware efficiency. 
%For instance, by sharing adder and comparator, the $processing~unit$ (PU) is compatible with both regular polar codes and fast constituent polar codes.      

%\begin{table}[h]
%\setlength{\extrarowheight}{2pt}
%\small
%\centering
%\caption{Number of all constituent codes with different sizes in a (1024, 512) polar code with rate of $0.5$}
%\begin{tabular}{cccccccc}
%\Xhline{1.2pt}
%\multirow{2}{*}{} & \multicolumn{6}{c}{Constituent codes sizes} & \multirow{2}{*}{All} \\ \cline{2-7}
%                  & 4     & 8    & 16   & 32   & 64   & 128   &                      \\ 
%\Xhline{1.2pt}
%$\mathcal{N}^{0}$                & 3     & 3    & 2    & 2    & 0    & 1     & 11                   \\
%$\mathcal{N}^{1}$                 & 3     & 3    & 2    & 1    & 0    & 0     & 9                    \\
%$\mathcal{N}^{REP}$                 & 16    & 8    & 4    & 1    & 1    & 1     & 31                   \\
%$\mathcal{N}^{SPC}$                 & 15    & 5    & 3    & 1    & 1    & 0     & 25                   \\ 
%\Xhline{1.2pt}
%\end{tabular}
%\label{codes_sizes}
%\end{table}

\subsection{System Overview}
\label{System Overview}
\begin{figure}[!t]
\centering
\includegraphics[width=2.5in]{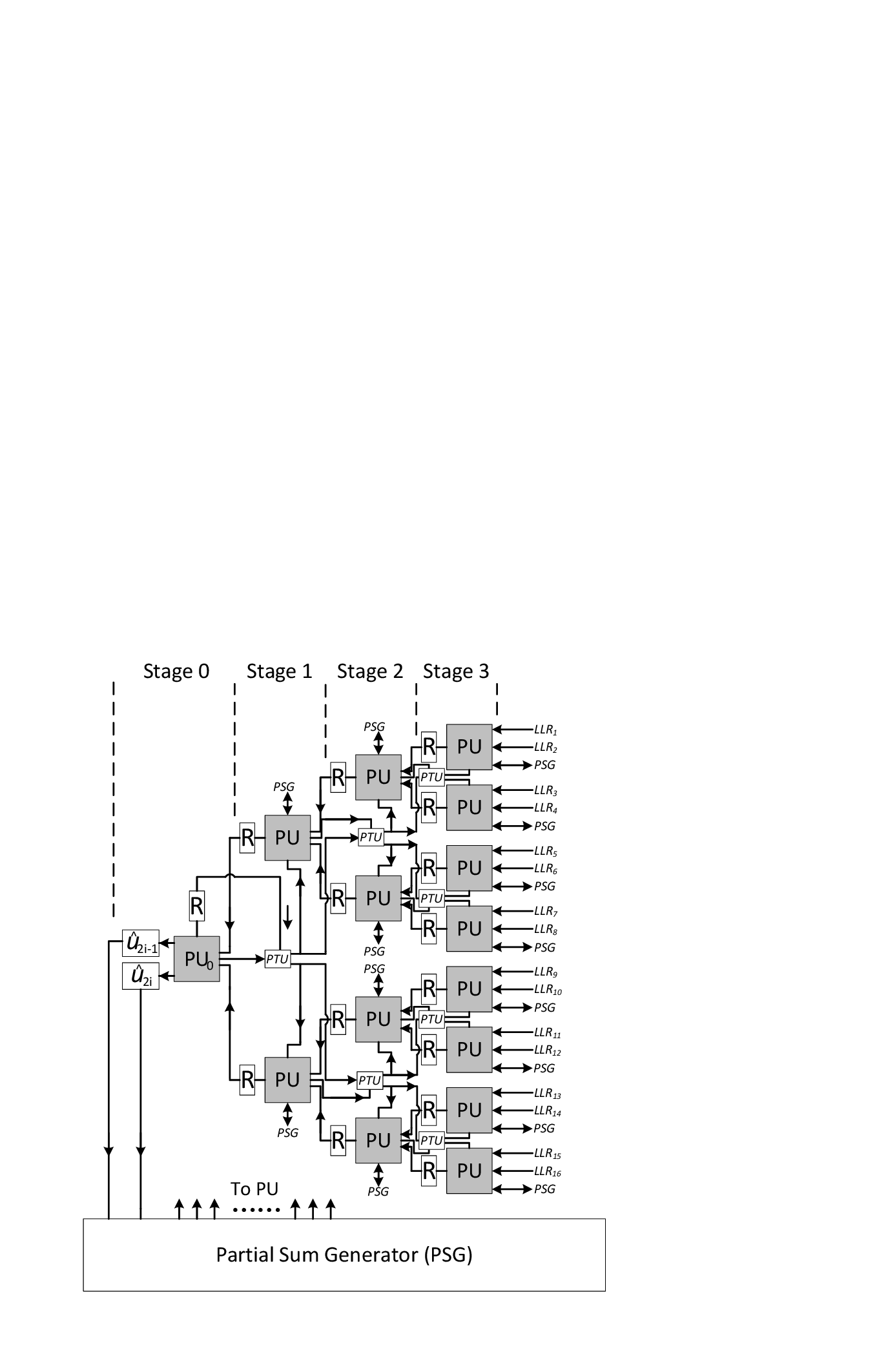}
\caption{Overview of proposed system when code length = 16}
\label{ststem_overview}
\end{figure}

As introduced in~\cite{leroux2013semi}, tree architecture or line architecture for SC decoder is the most common. 
Line architecture has a higher hardware utilization but needs increased complexity in control module and memory access. Thus, we adopt tree architecture in our design. 
Fig.~\ref{ststem_overview} shows an overview of proposed system when code length = 16. 
$Processing~unit$ (PU) performs the $f$ and $g$ functions in Eq.~(\ref{left_child}) and Eq.~(\ref{right_child}), respectively, and its arithmetic part is used to decode $\mathcal{N}^{SPC}$ and $\mathcal{N}^{REP}$ as well. 
Pre-computation technique is also used, which allows the $f$ and $g$ functions update in the same clock cycle. 
The PU used in stage 0 has a slight difference with ordinary PU. We denote it with PU$_0$ in the figure. 
According to Eq.~(\ref{min_SPC}), the minimum LLR value needs to be found. The comparator tree is used to perform this since it inherently exists in the tree architecture of PUs. A judicious scheduling permits obtaining the minimum value at $stage~0$ and recording the choice of smaller input for each PU at each stage. After that, a backward operation implemented by a series of $parity~transmit~unit$ (PTU) can help to locate the minimum one among the length $N$ $\mathcal{N}^{SPC}$ constituent polar codes. 
Design details are illustrated in section~\ref{Processing Unit Design}. 
The estimation of current bit in SC decoding is bases on the information of previous decoded bits ($\bm{\beta}$). This information is also called partial sum. 
Thus, a $partial~sum~generator$ (PSG) which can co-operate with decoding pipeline is also needed. 
We adopt the PSG introduced in~\cite{zhang2013low} in our design, and it is compatible with our system. Thus, the design of PSG is not discussed in this paper.          

\subsection{Dataflow, latency and flexibility analysis }
\label{Scheduling of Fast SC Decoder}

In terms of tree presentation, SC decoder conventionally process one node in each clock cycle. Traversal of a subtree contained $N$ leaf nodes needs $2N-2$ clock cycles. By using pre-computation as introduced in~\cite{zhang2012reduced}, which calculate the $f$ function and all the possible result of $g$ functions in the same clock cycle, the latency can be reduced to $N-1$. In our design, if this subtree is belong to fast constituent polar codes, the latency can be further reduced. 

For $\mathcal{N}^{0}$, the $\bm{\beta_v}$ are all set to $0$, and for $\mathcal{N}^{1}$, the $\bm{\beta_v}$ are determined by hard decision of input LLRs. Both of the two computations need only one clock cycle after they are activated. 
For $\mathcal{N}^{SPC}$, according to Eq.~(\ref{HD_SPC}), Eq.~(\ref{min_SPC}), and Eq.~(\ref{beta_SPC}), only three operations needed. Finding the minimum LLR can be done by a comparator tree, which is naturally existed in SC decoder with tree architecture since every PU has a comparator for Eq.~(\ref{left_child}). 
For $N$ LLRs, finding the smallest one use $Log_2N$ clock cycles. Meanwhile, we can obtain the parity bit when the minimum LLR is found, which will be explained in the next subsection. After that, one more clock cycle is need for signal parity check which is done by a $XOR$ gate. Thus, totally, decoding a length $N$ $\mathcal{N}^{SPC}$ constituent polar codes need $Log_2N + 1$ clock cycles. 
For $\mathcal{N}^{REP}$, according to Eq.~(\ref{beta_REP}), an accumulation operation is needed. Similar to the comparator tree, an adder tree also exists in SC decoder within the tree architecture since every PU has an adder for Eq.~(\ref{right_child}).  For a length $N$ $\mathcal{N}^{REP}$ constituent polar code, it needs $Log_2N$ clock cycles to decode.
    
%\begin{table}[h]
%\setlength{\extrarowheight}{1pt}
%\centering
%\caption{Summary of decoding latency for each constituent polar code}
%\begin{tabular}{|c|c|c|c|c|}
%\hline
%Type      &	$\mathcal{N}^0$	&	$\mathcal{N}^1$	&	$\mathcal{N}^{SPC}$	&	$\mathcal{N}^{REP}$     \\ \hline
%Latency(clock cycle)	&	1	&	1	&	$Log_2N + 1$	&	$Log_2N$             \\ \hline
%\end{tabular}
%\label{latency_summariy}
%\end{table}

%Table~\ref{latency_summariy} gives the summary of decoding latency for each constituent polar code. 
$\mathcal{N}^0$ and $\mathcal{N}^1$ have time complexity $O(1)$ and 	 $\mathcal{N}^{SPC}$ and $\mathcal{N}^{REP}$ have time complexity $O(log_2N)$. 
Compared with commonly discussed SC architecture in~\cite{leroux2013semi},~\cite{zhang2012reduced} and~\cite{yuan2014low}, which all have linear time complexity $O(N)$, we can benefit significantly from proposed scheduling scheme in term of latency, especially with very large $N$. The latency reduction of $N=1024$ polar code with different rate will be presented in the next section.  

The main challenge for fast-SSC decoder is that the architecture subject to the rate of codes. This is due to the reason that polar codes with different rates do not have the uniform distribution of constituent polar codes. Proposed design overcomes this obstacle by exploring the similarity between the decoding architecture of fast constituent and regular polar codes. The specific designed PU allows the tree architecture to deal with both fast constituent and regular polar codes, which means the entire decoding processing can run smoothly no matter what the distributions of constituent codes are. This architecture is independent and does not relay on the distribution of constituent codes. This property provides the flexibility for multiple rates. To switch from one rate to another rate, only the control signals for given PUs need to be modified.   

\subsection{Processing Unit Design}
\label{Processing Unit Design}
\begin{figure}[!t]
\centering
\subfloat[]{\includegraphics[width=2.5in]{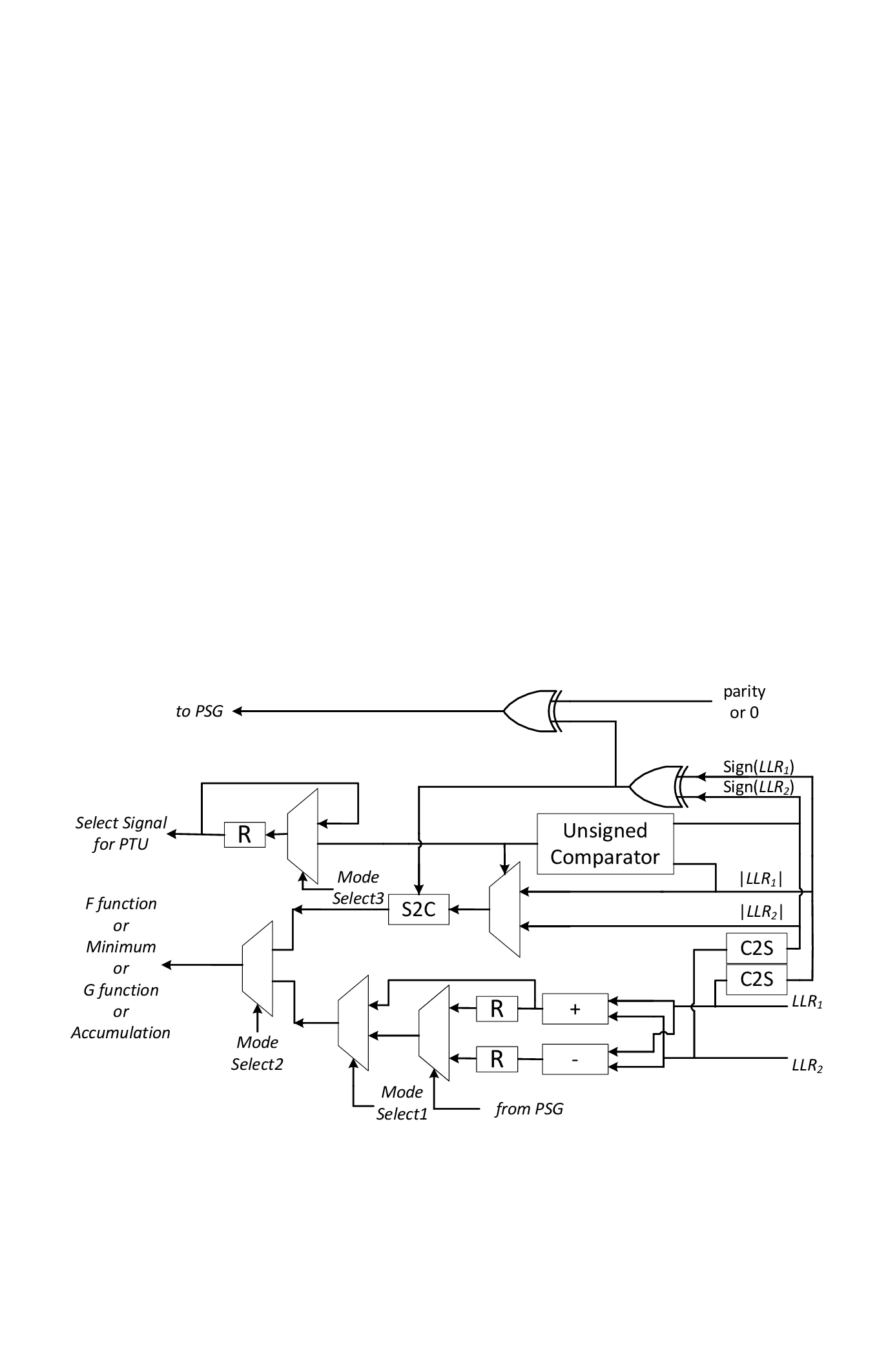}\label{processing_unit}}
\vfil
\subfloat[]{\includegraphics[width=2.5in]{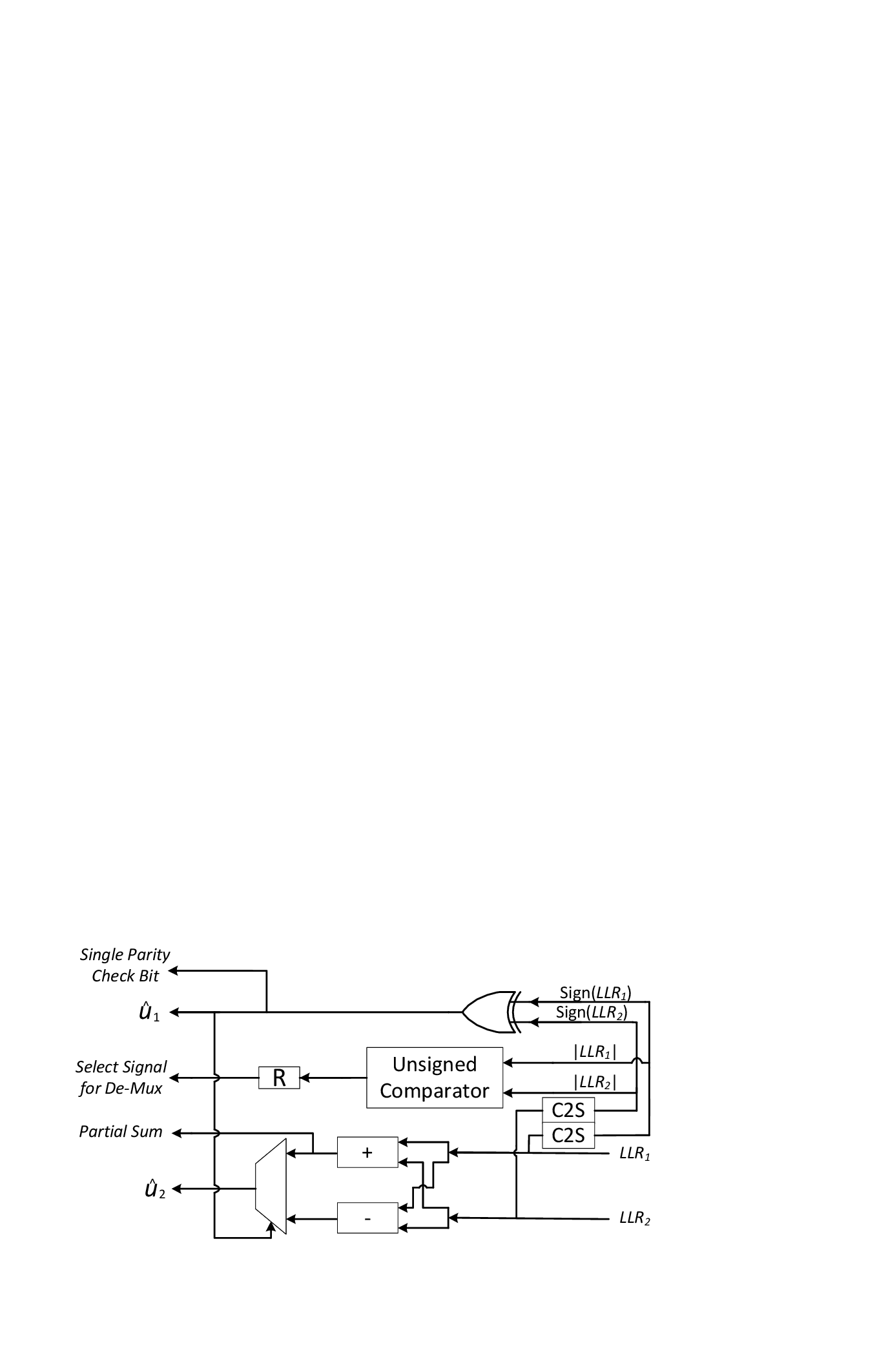}\label{processing_unit0}}
\caption{\protect\subref{processing_unit} Design details of PU, \protect\subref{processing_unit0} Design details of $PU_0$}
\label{processing unit design}
\end{figure}  
Fig.~\ref{processing_unit} shows design details of PU. A single PU can perform $f$ and $g$ functions in Eq.~(\ref{left_child}) and Eq.~(\ref{right_child}), respectively. Also a PU tree can help to find the minimum values or do accumulation for multiple inputs. In Fig.~\ref{processing_unit}, $S$ stands for $signed~magnitude~number$ and $C$ stands for $2's~ complement~number$. 
Unlike the PU design in~\cite{yuan2014low}, in which data are initially stored as signed magnitude form, our design use 2's complement as initial form. We do this for two reasons. 
1). According to synthesis result, the critical path of PU is along with the $g$ function path. By moving number system convert modules to the $f$ function path, which means using 2's complement as initial data form, the critical path is still along with $g$ function path, but with significant reduction. 
2). Compared with four number system convert modules are used in~\cite{yuan2014low}, only three are used if use 2's complement number. This is more hardware efficient. 
The benefits of this modification can be seen in section \ref{Hardware Analysis of Comparison}.    

For each PU, two LLRs are fed simultaneously. Since we use the pre-computation technique, $f$ and $g$ functions are calculated at the same time, and which one needs to be output is determined by $mode~select~2$. According to Eq.~(\ref{right_child}), there are only two types of possible results for $g$ function, sum or difference. Its final result depends on the corresponding partial sum. So two registers are used here to hold the most recently computed values until the corresponding partial sum is calculated. When it calculates the sum for decoding $\mathcal{N}^{REP}$, only additions are needed. The datapath is decided by $Mode~select~1$ signal. When $f$ function is performed, according to Eq.~(\ref{left_child}), both 2 inputs are divided into two parts: sign bit and unsigned number. Each part is processed separately first, and then results of two parts are combined together to obtain the updated value. $C~to~S$ and $S~to~C$ modules are needed before and after comparisons, respectively.  
When it deals with $\mathcal{N}^{SPC}$, the result of comparison should be recorded using a register as the $select~signal$ for PTU. 
Since the processing of searching minimum value lasts several clock cycles, there should be a feedback of the register to hold this value for the later clock cycles. The input source is chosen by $Mode~select~3$ signal.   
Since every PU does $exclusive~or~operation$ to the sign bit of two inputs, according to Eq.~(\ref{HD_SPC}), the sign bit of the final value in stage 0 should be equal to the parity. 
%%If current PU received confirm signal from De-Mux to indicate itself contains the minimum value, an $exclusive operation$ is executed with current hard-decision and parity then output the result as partial sum for these stage. Otherwise, no need do any parity check.
Eq.~(\ref{beta_SPC}) can be performed using an $XOR$ gate. The PU that contains the minimum LLR receives the parity check bit and the others receive $0$s. The transmission of parity check bit is done by the PTU which is a two input two output module. One input is the $parity~check~bit$ (PCB) and the other is the $select~signal$ (SS). The parity check bit is transmitted via $output~1$ (O1) or $output~2$ (O2) bases on the values of SS. Table.~\ref{truth table} shows the truth table of PTU. We can obtain the logic expression of O1 and O2 as: $O1~=~PCB~and~\overline{SS}~,O2~=~PCB~and~SS$. This can be done by two $and$ gates and one $Inverter$.
\begin{table}[h]
\centering
\caption{Truth table of PTU}
\begin{tabular}{|c|c|c|c|c|c|c|c|}
\hline
PCB & SS & O1 & O2 & PCB & SS & O1 & O2 \\ \hline
0   & 0  & 0  & 0 & 1   & 0  & 1  & 0 \\ \hline
0   & 1  & 0  & 0 & 1   & 0  & 0  & 1 \\ \hline

\end{tabular}
\label{truth table}
\end{table}

The PU in $stage 0$, as denote PU$_0$ in~Fig.~\ref{ststem_overview}, has a simpler architecture. Fig.~\ref{processing_unit0} shows the design details of $PU_0$. Since only one more clock cycle need for single parity check, there is no feed back to this register. Furthermore, $\mathcal{N}^{SPC}$ cannot exist in $stage 0$. So top part in Fig.~\ref{processing_unit} which is relative to single parity check can be removed. For $g$ function and $\mathcal{N}^{REP}$ , the output of $f$ function can be feed back to it immediately, and the sign bit of the result of adding is the partial sum for $\mathcal{N}^{REP}$.       
      
%
%\begin{figure}[!t]
%\centering
%\includegraphics[width=3 in]{processing_unit.eps}
%\caption{Design details of PU}
%\label{processing_unit}
%\end{figure}
%
%\begin{figure}[!t]
%\centering
%\includegraphics[width=2.5 in]{processing_unit0.eps}
%\caption{Design details of $PU_0$}
%\label{processing_unit0}
%\end{figure}

%\begin{table*}[!t]
%\centering
%\caption{Hardware comparison of different $(n,k)$ SC decoder with $q$-bit quantization for inner LLRs using tree architecture}
%\begin{tabular}{|c|c|c|c|c|}
%\hline
%Hardware Type           & SC-Precomputation~\cite{zhang2012reduced} 	& SC~\cite{leroux2011hardware} 					& 2b-SC-Precomputation~\cite{yuan2014low} 	& Proposed Design \\ \hline
%\# of PE                &  $n-1$                 		&   $n-1$  						&     $n-2$(and $+~1~p$ node)	&      $n-1$       \\ \hline
%\# of PTU               &  $0$                 			&   $0$ 						&     $0$               		&      $2/n-1$          \\ \hline
%\# of 1 bit REG         &  $\thickapprox 3qn$           &   $\thickapprox qn$ 			&     $\thickapprox 3qn$        &      $\thickapprox (3q+1)n$           \\ \hline
%Latency (clock  cycle)  &  $n-1$                 		&   $2n-2$ 						&     $0.75n-1$             	&      $\thickapprox (0.3\thicksim 0.1)n$ (at rate between 0.05 and 0.95)           \\ \hline
%Throughput (Normalized) &  $2$               			&   $1$ 						&     $2.67$         			&      $\thickapprox6.69\thicksim 22.26 $ (at rate between 0.05 and 0.95)        \\ \hline
%\end{tabular}
%\label{comparison}
%\end{table*}
\begin{table}[!t]
\centering
\caption{Hardware comparison of different $(n,k)$ SC decoder with $q$-bit quantization for inner LLRs using tree architecture}
\begin{tabular}{|c|c|c|c|c|}
\hline
Hardware Type           & \cite{zhang2012reduced} 	& \cite{leroux2011hardware} 					& \cite{yuan2014low} 	& Proposed Design \\ \hline
\# of PU                &  $n-1$                 		&   $n-1$  						&     $n-1$	&      $n-1$       \\ \hline
\# of PTU               &  $0$                 			&   $0$ 						&     $0$               		&      $2/n-1$          \\ \hline
\# of 1 bit REG         &  $\thickapprox 3qn$           &   $\thickapprox qn$ 			&     $\thickapprox 3qn$        &      $\thickapprox (3q+1)n$           \\ \hline
HC 	&	$1.3$	&	$1$&	$1.3$&$1.31$ \\ \hline
Latency (clock  cycle)  &  $n-1$                 		&   $2n-2$ 						&     $0.75n-1$             	&      $\thickapprox (0.1\thicksim 0.3)n$            \\ \hline
Throughput  &  $2$               			&   $1$ 						&     $2.67$         			&      $\thickapprox6.69\thicksim 22.26 $        \\ \hline
Throughput/HC	&	$1.53$ &	$1$ &	$1.74$&$5.1\thicksim 16.99$ \\ \hline
\end{tabular}
\label{comparison}
\end{table}

\subsection{Fixed point analysis}
\label{Quantization Analysis}

\begin{figure}[!t]
\centering
\includegraphics[width=3 in]{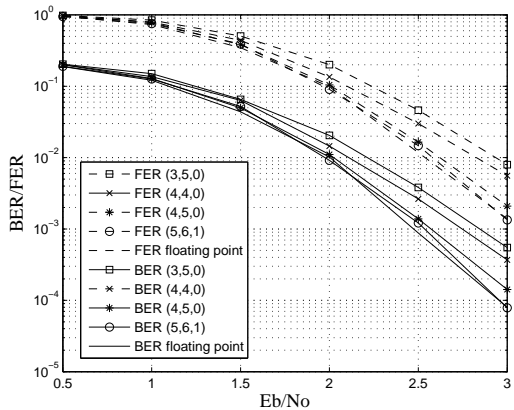}
\caption{Effect of quantization on the BER/FER performance of $(1024,512)$ code}
\label{fix_point}
\end{figure}

Fig.~\ref{fix_point} shows the effect of quantization on the $(1024,512)$ polar code. For channel outputs and inner LLRs, we use separate quantization schemes. 
The quantization schemes are shown in $(C,L,F)$ format. 
%Where $C$ is the total number of bits used for presenting channel output, $L$ is the total number of bits used for presenting inner LLRs and $F$ is the total number of bits used for fraction parts of both channel output and LLRs.
Where $C$, $L$ and $F$ are the number of bits used for presenting channel output, inner LLRs and fraction parts of both channel output and LLRs, respectively.
Since no multiplication or division used, which means the length of fraction does not change, channel outputs and inner LLRs use the same fraction precision.  
As the result of the trade-off between hardware efficiency and decoding performance, we choose $(4,5,0)$ quantization scheme in our design.  
%we can observe that there are only about $0.1dB$ performance lost with $(4,5,0)$ scheme, and increasing $1$ bit fraction precision based on that, namely $(5,6,1)$, yields only less than $0.1dB$ increased on performance. However, if we try to reduce the precision on the channel output or inner LLRs by $1$ bit, namely $(3,5,0)$ or $(4,4,0)$, respectively, they cause $0.4dB$ and $0.3dB$ performance loss, respectively. From these results above, 
 
\section{Hardware Analysis and Comparison}
\label{Hardware Analysis of Comparison}

\begin{figure}[!t]
\centering
\includegraphics[width=3in]{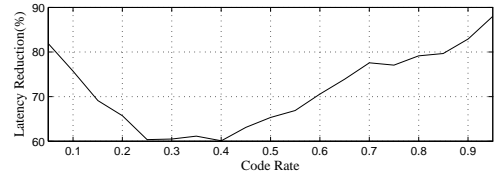}
\caption{Latency Reduction $vs.$ Code Rate}
\label{latency_reduction}
\end{figure}

In this section, the comparisons between proposed design and other state-of-the-art designs are given, and synthesis results using $Nangate~FreePDK~45nm$ process are also presented. Table.~\ref{comparison} shows the hardware comparison of different $(n,k)$ SC decoders with $q$-bit quantization for inner LLRs using tree architectures. All the throughputs and hardware complexity (HC) are normalized to the SC decoder in~\cite{leroux2011hardware}, and the hardware complexity is estimated based on the synthesis results. The latency for proposed design is a range with respect to the code rates change from $0.05$ to $0.95$. From this table, we can see that our proposed design achieves the highest throughput per unit of hardware complexity. The exact latency depends on the code rate. Fig.~\ref{latency_reduction} shows the latency reduction of the proposed design along with code rates from $0.05$ to $0.95$. The reduction is relative to the 2b-SC-Precomputation decoder which so far is known to be the fastest. 
The figure shows at least $60\%$ latency reduction can be achieved by our proposed design.
%The worst point is at the rate of about $0.4$. Beyond that, the reduction increases along with the code rate. 
This is very promising for many applications where high rate channel codes are needed, such as for data storage system.  
%In order to achieve low latency, some extra hardware overhead cannot be avoided. 
%This latency reduction is achieved with minimum hardware overhead.
%In our proposed design, additional $n$ registers and $n/2-1$ PTUs are needed. Since the PTU only consists of two $AND$ gate and one $Inverter$, and only one more register which is for $selecte~signal$ is added to the original $3q$ registers for each PU, this overhead is negligible. 

Additionally, we implemented the proposed design with $Verilog$ for the polar code with length=$1024$ and synthesized it using $Nangate~FreePDK~45nm$ process with $Synopsys~Design~Complier$. We calculated the throughput for $(1024,870)$ and $(1024,512)$ polar codes. Table~\ref{syn_result} shows the synthesis result for $(1024,870)$ and $(1024,512)$ polar codes. Notice that the maximum frequency is higher than that reported in \cite{yuan2014low} which use the same process as our design. Our design in theory should have a lower maximum frequency since we have one more Mux delay for regular and fast constituent polar codes. This performance improving is attributable to the modification we have done to PU as described in section \ref{Processing Unit Design}.

\begin{table}[h]
\centering
\caption{Synthesis result for $(1024,870)$ and $(1024,512)$ polar codes}
\begin{tabular}{|c|c|}
\hline
Silicon Area ($\mu m^2$)                                     & 275899                  \\ \hline
Max Frequency (GHz)                                  & 1.04                    \\ \hline
Latency (1024,870) (clock cycle)					    & 156					 \\ \hline
Throughtput(1024,870) (Gbps)                        & 5.81                    \\ \hline
Latency (1024,512) (clock cycle)                     & 266                     \\ \hline
Throughtput(1024,512) (Gbps)                        & 2.01   \\ \hline                     
\end{tabular}
\label{syn_result}
\end{table}

\section{Conclusion}
\label{Conclusion}

In this paper, we proposed a hardware architecture of fast-SSC algorithm for polar codes. By exploiting the similarity between the decoding processing of fast constituent and regular polar codes, proposed design overcomes the disadvantage of fast-SSC decoder that lacking decoding flexibility with respect to multiple code rates.   Corresponding scheduling plan and the intendedly designed PU are also described. Result shows that proposed design significantly increase the decoding throughput of polar codes compared with other state-of-art SC decoders.

\bibliographystyle{IEEEtran}
\bibliography{IEEEabrv}

\end{document}